\documentclass[prb,reprint]{revtex4-2}
\usepackage{amsmath}
\usepackage{bm} 
\usepackage{amssymb}
\usepackage{titlesec}
\usepackage{color}
\usepackage{multirow}
\usepackage{url}
\usepackage[colorlinks,bookmarks=false,citecolor=red,linkcolor=blue,urlcolor=blue]{hyperref}
\usepackage{graphicx}
\usepackage{mathrsfs,bbm}
\usepackage{wrapfig}
\usepackage{physics}
\usepackage{enumerate}
\usepackage[version=3]{mhchem}
\usepackage{gensymb}
\usepackage{bm}
\usepackage{xcolor}

\usepackage{comment}

\def\chat{\hat{c}}
\def\nhat{\hat{n}}
\def\Hhat{\hat{H}}
\def\fhat{\hat{f}}

\begin{document}
 
\title{Evolution of quasiparticle edge states with Hubbard interaction in Rice-Mele chain}

	\author{Jyoti Bisht}
	\email{jyotibisht9876@gmail.com}
	\affiliation{School of Physical Sciences, Jawaharlal Nehru University, New Delhi 110067, India.}
	\author{Brijesh Kumar}
	\email{bkumar@mail.jnu.ac.in}
	\affiliation{School of Physical Sciences, Jawaharlal Nehru University, New Delhi 110067, India.}
	\date{\today}
	
\begin{abstract}
	We study the behaviour of edge states in Rice-Mele model with Hubbard interaction, $U$, at half-filling using density matrix renormalization group, exact diagonalization and effective charge dynamics in Kumar representation. For a fixed dimerization, $\delta$, and staggered potential, $V$, we find by increasing $U$ the quasiparticle edge states in the charge gap to come down in energy from $V$ in the absence of Hubbard interaction to zero energy for $U\approx2V$. This presents an uncommon case where repulsion leads to zero-energy edge states. Upon increasing $U$ further, the edge state energy starts increasing again until they are lost in the bulk. However, upon increasing $U$ even further, these edge states reappear in the high energy gap. So, with Hubbard interaction, the edge states in Rice-Mele chain transmigrate from the physical charge gap to a high energy gap.
\end{abstract}

\maketitle

\section{Introduction}

The Rice-Mele model is an interesting problem of one-dimensional electrons, conceived as a generalization of the Su-Schrieffer-Heeger (SSH) model to diatomic polymers~\cite{rice1982elementary,su1979solitons}. In its simplest form, it is a non-interacting tight-binding model with dimerized nearest-neighbour hopping (caused by static Peierls distortion~\cite{Peierls.book}) and a staggered onsite potential. At half-filling, the Rice-Mele model presents an example of a topological insulator with edge states. It has been quite fruitful in studying electric polarization and charge pumping~\cite{Asboth2016,cayssol2021topological,fuchs2021orbital}, and also realized in optical lattices for its topological properties~\cite{atala2013direct,przysikezna2015rice,biedron2016topological,cooper2019topological}. 

In the absence of staggered potential, i.e. in the SSH model, we recently investigated the behaviour of edge states with interaction, and discovered that, by increasing the interaction, the edge states transmigrate from the charge gap to a high-energy gap via an intermediate phase with no edge states~\cite{bisht2024transmigration}. How this transmigration of edge states would take effect in the presence of staggered potential is not clear and has not been investigated before. Previous works on Rice-Mele model with Hubbard interaction have explored different aspects of polarization~\cite{requist2018accurate,morimoto2021electric,aligia2023topological,segura2023charge} and Thouless pumping~\cite{nakagawa2018breakdown,yang2024nonadiabatic,walter2023quantization,viebahn2024interactions} for correlated electrons, but a basic understanding of the evolution of edge states with interaction is still found lacking. Therefore, in this paper, we study the half-filled Rice-Mele-Hubbard (RMH) model for its edge state behaviour, as we did for the interacting SSH chain in Ref.~\cite{bisht2024transmigration}. 

We perform this study by employing a combination of numerical and analytical methods involving density matrix renormalization group (DMRG), exact diagonalization (ED), and effective charge dynamics in Kumar  representation~\cite{bisht2024transmigration,kumar2008canonical}. In Sec.~\ref{sec:RMHubbard_DMRG}, we carry out a purely numerical investigation of the edge states (arising due to dimerization, $\delta$) of the charge quasiparticles in the RMH chain using DMRG and ED methods~\cite{itensor}. For a fixed staggered potential, $V$, by increasing the Hubbard interaction, $U$, we find the edge state energy to vary non-monotonously. First it decreases with $U$ upto a critical value, $U_c \approx 2V$, where the edge state energy actually becomes zero! Thereafter, it starts increasing with $U$, and the edge states undergo transmigration from the physical charge gap to the high energy gap for strong correlations. Notably, for any strong enough $\delta$, we always find a $U_c$ where the edge states invariably occur at zero energy. This is a remarkable happenstance of competition between $V$ and $U$. Generally, for non-zero values of $V$ or $U$, the quasiparticle edge states are found to have non-zero energies. So, it is quite a surprise to find them at zero energy inspite of the interaction. However, on a careful reflection, the existence of these zero energy edge states looks inevitable, as it marks the change from band to Mott insulator across $U_c$. In Sec.~\ref{sec:RMHubbard_Theory} we present a theory capable of describing the physics of quasiparticle edge states in the RMH chain. The results obtained from this theory are compared with those obtained from DMRG and ED. We find a good qualitative match between them, and a strong agreement of the edge energies for weak to moderate correlations. A phase diagram in the $U$-$V$ plane is also worked out for a fixed dimerization ($\delta=0.7$) from this theory to highlight the key features of the edge state behaviour resulting from the competition between the staggered potential and the Hubbard interaction. We end this paper with a summary in Sec.~\ref{sec:summary}.

\section{\label{sec:RMHubbard_DMRG} Edge States in Rice-Mele-Hubbard Chain}
The Rice-Mele-Hubbard Hamiltonian on an open chain of $L$ sites can be written as follows:
\begin{equation}
 \begin{split}	
	\Hhat &= -t\sum_{l=1}^{L-1} \sum_{s=\uparrow,\downarrow} \left[1 + (-)^{l}\delta \right] \left(\chat^\dagger_{l,s}\chat^{}_{l+1,s} + {\rm H.c.}\right) \\
	&+ V\sum_{l=1}^{L}\sum_{s=\uparrow,\downarrow} (-)^l \nhat_{l,s} + U\sum_{l=1}^L\left(\hat{n}^{ }_{l,\uparrow} - \frac{1}{2}\right)\left(\hat{n}^{ }_{l,\downarrow} - \frac{1}{2}\right)
 \end{split}  \label{eq:RMHubbard}	
\end{equation}  
where the first term is the SSH model with $\delta$ as the dimerization parameter and $t$ as the nearest neighbour hopping amplitude, $V$ is the staggered local potential, and $U$ represents the Hubbard interaction. 

We study the edge state behaviour of RMH model by investigating Eq.~\eqref{eq:RMHubbard} for different values of dimerization ($\delta = 0.5$, $0.7$) and staggered potential ($V=1$, $2$, $5$). We put $t=1$ in the calculations, and employ DMRG method to compute the edge state energy ($\varepsilon_1$) and the bulk charge gap ($\Delta_c$). The energy of adding a particle to the half-filled system, i.e. $ E_g(L+1)- E_g(L)$, is $\varepsilon_1$ for $\delta>0$, and $\Delta_c$ for $\delta<0$. Here $E_g(N_e)$ denotes the ground state energy of the RMH chain for $N_e$ electrons; note that $N_e=L$ corresponds to half-filling. The system size used to calculate $\varepsilon_1$ and $\Delta_c$ is $60$, with $200$-$400$ and $500$-$1500$ sweeps respectively, while maintaining the maximum truncation error $\approx 10^{-12}$.

\begin{figure}[t]
	\begin{center}
		\includegraphics[width=\columnwidth]{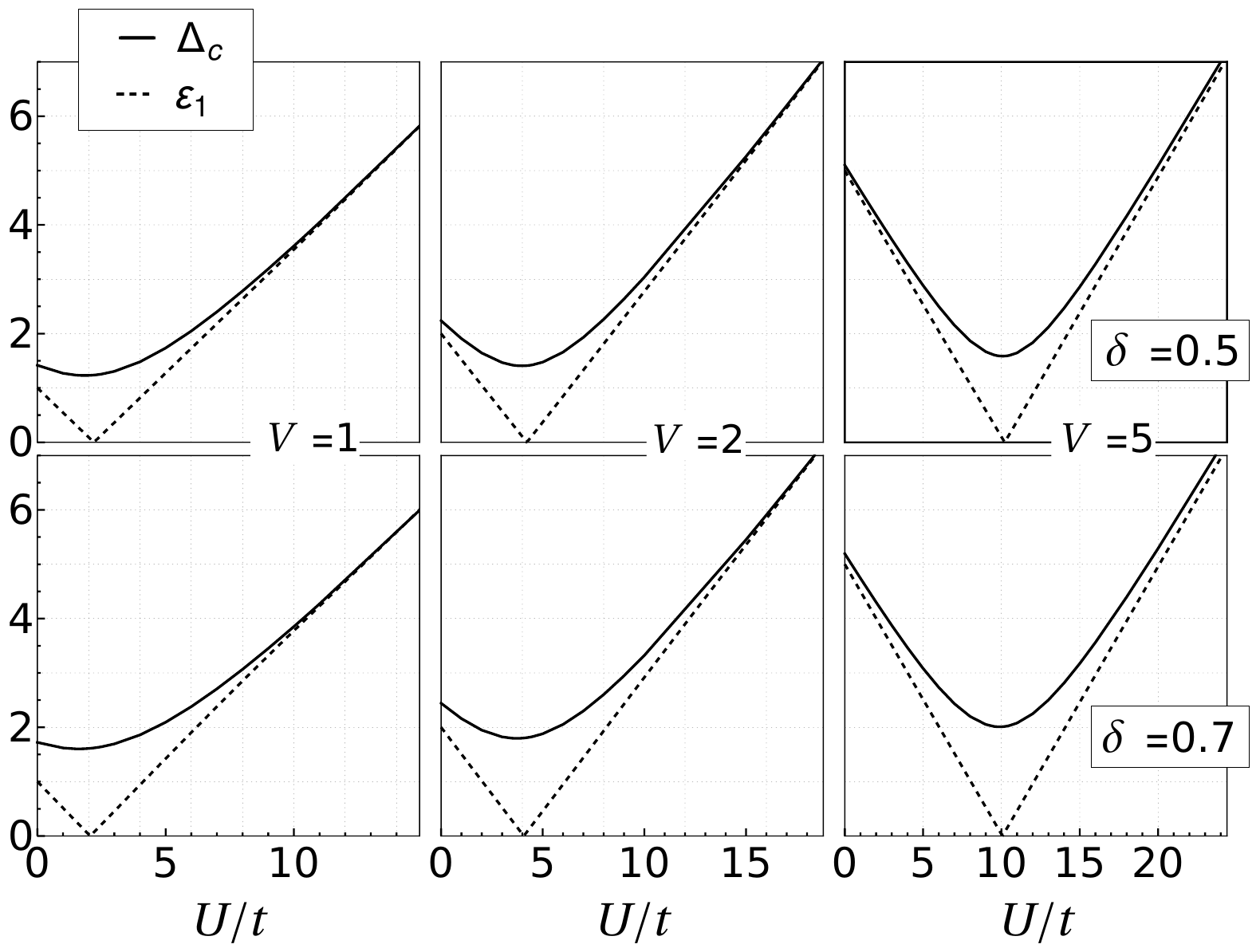}
		\caption{Evolution of the bulk charge gap, $\Delta_c$, and the edge state energy, $\varepsilon_1$, with Hubbard interaction, $U$, in units of hopping, $t$, for different values of staggered potential, $V$, and dimerization, $\delta$, in the Rice-Mele-Hubbard chain at half-filling. Notably, the cone-shaped $\varepsilon_1$ vs. $U$ touches zero energy on its tip at the critical interaction, $U_c \approx 2V$, and for a large enough $U$, the edge state merges with the bulk.}
		\label{fig:delta=0.5_0.7_edgeEn_BGap}
	\end{center}
\end{figure}

Let us check the limiting case, $U=0$, first. The non-interacting Rice-Mele chain has the bulk charge gap, $\sqrt{V^2 + 4\delta^2 t^2}$, and the edge state energy, $V$. Indeed, we get their correct values numerically for $U=0$, as shown in Fig.~\ref{fig:delta=0.5_0.7_edgeEn_BGap}. For $U\neq0$, i.e. in the RMH chain, the edge state energy is found to show a prominent cone-like variation with $U$. Notably, the tip of this cone touches the zero energy, and the `critical' value of the Hubbard interaction at this tip, $U_{c}$, is directly proportional to the value of the staggered potential. We find that $U_c \approx 2V$ does not seem to depend strongly on the degree of dimerization (here, we have $\delta=0.5$ and $\delta=0.7$).

As the interaction crosses $U_{c}$, the edge state energy increases gradually, and merges into the bulk at some point, $U_{c,1}$. It means, the edge states disappear beyond $U_{c,1}$, like in the SSH-Hubbard chain~\cite{bisht2024transmigration}. From Fig.~\ref{fig:delta=0.5_0.7_edgeEn_BGap}, it is clear that $U_{c,1}$ increases with increase in $V$ for a fixed $\delta$. Strong interaction is required for the edge state energy to overcome the bulk gap in the presence of strong staggered potential. For example, $U_{c,1}$ at $\delta=0.7$ is approximately $11$ and $17$ for $V=1$ and $2$ respectively.

\begin{figure}
	\begin{center}
		\includegraphics[width=\columnwidth]{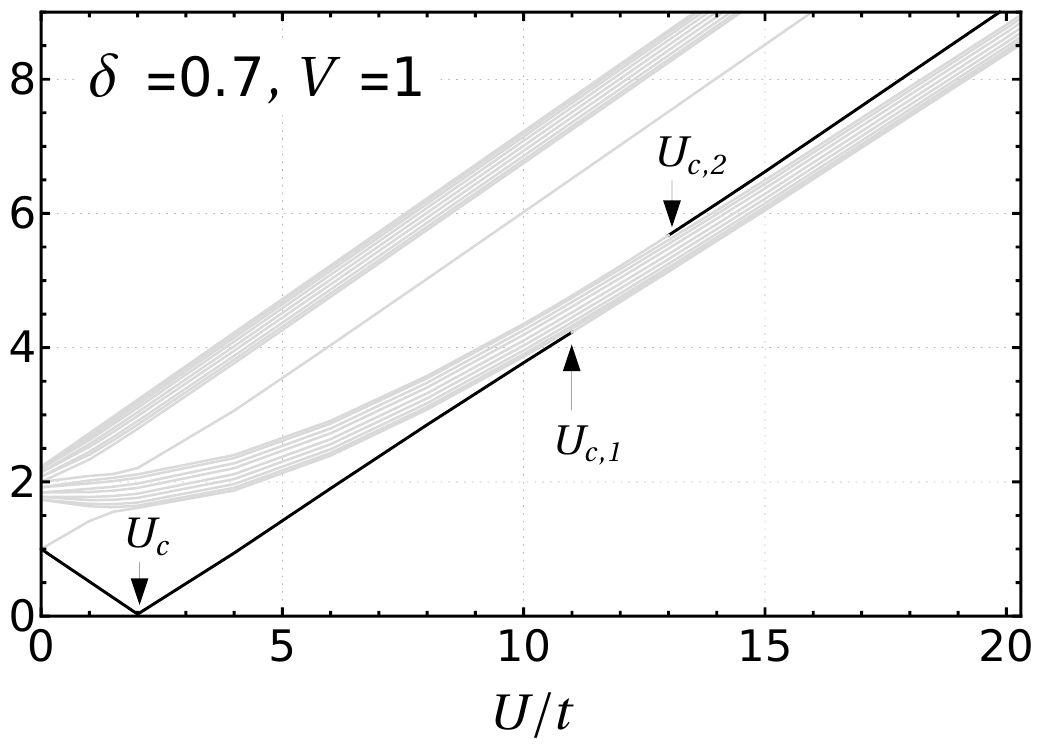}
		\caption{Energies of the successive quasiparticle excitations above the ground state of the half-filled RMH model, calculated using DMRG on an open chain of length, $L=20$, for $\delta=0.7$ and $V=1$. The lowest cone-shaped black line touching zero at $U_c \approx 2$ corresponds to the edge state in the charge gap; it is clearly seen to be lost into the bulk at $U_{c,1}\approx11$, and reappear in the high energy gap around $U_{c,2}\approx13$. }
		\label{fig:L20_DMRG_All}
	\end{center}
\end{figure}

Beyond $U_{c,1}$, the edge states stay lost upto a certain interaction strength $U_{c,2}$, after which they reappear in the high energy band gap! To see this transmigration of edge state from physical charge gap to high energy gap~\cite{bisht2024transmigration}, let us focus on Fig.~\ref{fig:L20_DMRG_All}, which for $\delta=0.7$ and $V=1$ shows the energy of quasiparticle excitations above half-filling. The  energy required to create the $N$th successive quasiparticle above the half-filled ground state is defined as $E(N) = \mathcal{E}(L+N) - \mathcal{E}(L+N-1)$, where $\mathcal{E}(L+N)=E_g(L+N)-E_g(L)$ is the minimal energy required to add a total of $N$ quasiparticles to the half-filled ground state. For instance, $E(1)$ is the energy, $\varepsilon_1$, of the edge state existing in the physical charge gap. The thing to observe here is the quasiparticle state emerging out of the bulk for $N=L/2$, i.e. $E(10)$ for $L=20$. For strong enough interaction, the edge state clearly reappears in the high energy gap. This transmigrated edge state is highlighted in Fig.~\ref{fig:L20_DMRG_All} by the thick black lines. 

\begin{figure}[htbp]
	\begin{center}
		\includegraphics[width=\columnwidth]{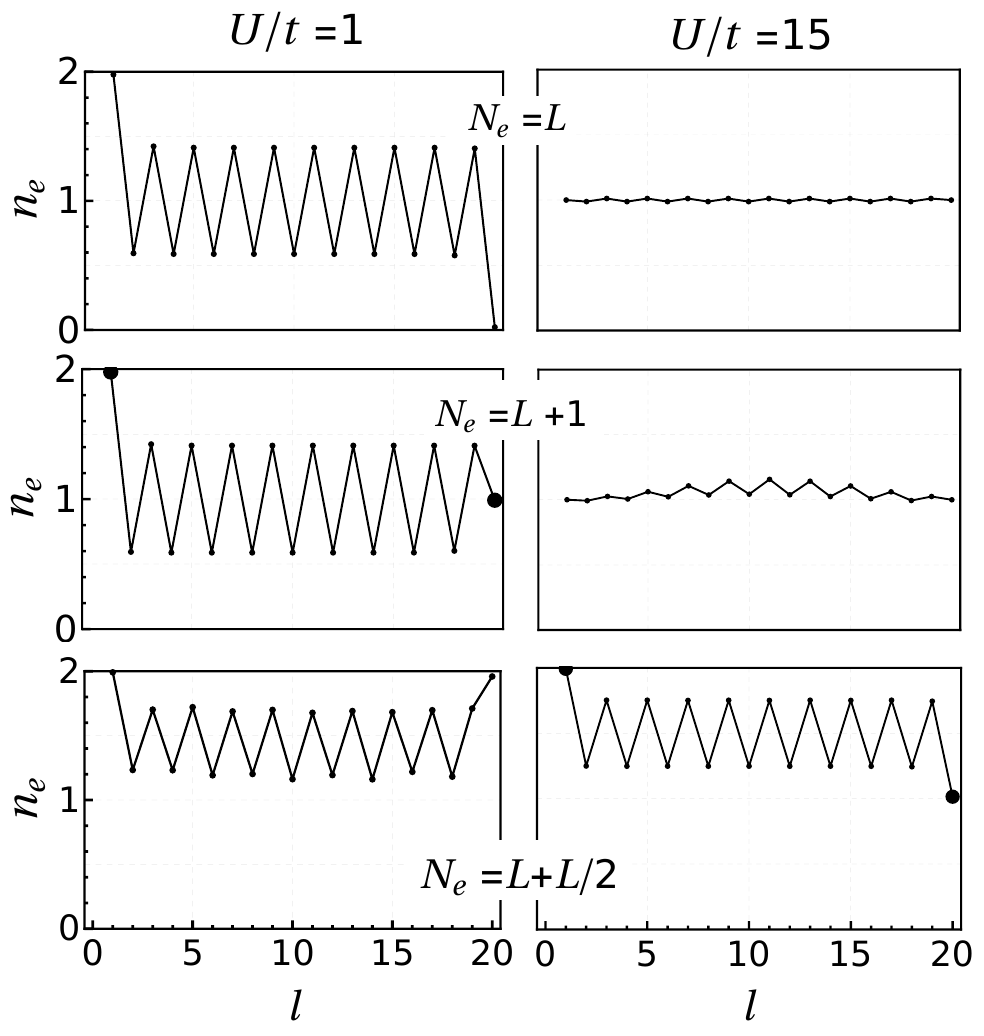}
		\caption{Tracking the edge state transmigration by calculating the expectation value of electron number, $n_e(l)$, on site, $l$, using DMRG for $L=20$. The middle panel for $U=1$ is for the quasiparticle edge state in the charge gap; notably, the same behaviour is exhibited in the lowest panel  for $U=15$, which is for the edge state in the high energy gap.}
		\label{fig:Edge_state_transmigration_d0p7_V1p0_new}
	\end{center}
\end{figure}

We also calculate the electron number expectation values, $n_{e}(l)=\langle \hat{n}_{l\uparrow} + \hat{n}_{l\downarrow} \rangle$, on every site in the ground state for three different electronic fillings, namely, the half-filling ($N_e=L$), one extra electron above the half-filling ($N_e=L+1$), and three-quarter filling ($N_e=L+L/2$) for $U/t=1$ and $U/t=15$. The $n_e(l)$ vs. $l$ for these three cases are plotted in Fig.~\ref{fig:Edge_state_transmigration_d0p7_V1p0_new}. In the ground state of the half-filled case, $n_e(l)$ at the edges is 2 or 0 for $U/t=1$, and 1 for $U/t=15$, as expected. For $U/t=1$, the edge state in the physical charge gap corresponds to $N_e=L+1$ case, wherein the electron occupancy at one end is changed to 1 (from the ground state occupancy, 0). This edge state transmigrates to the high energy gap for strong interactions. Hence, for $U/t=15$, we see the same behaviour at the edges in $n_e(l)$ vs. $l$ for $N_e=L+L/2$ as for $N_e=L+1$ when $U/t=1$. These edge states in Fig.~\ref{fig:Edge_state_transmigration_d0p7_V1p0_new} are marked by two big dots at the two ends.

\begin{figure}[htbp]
	\begin{center}
		\includegraphics[width=\columnwidth]{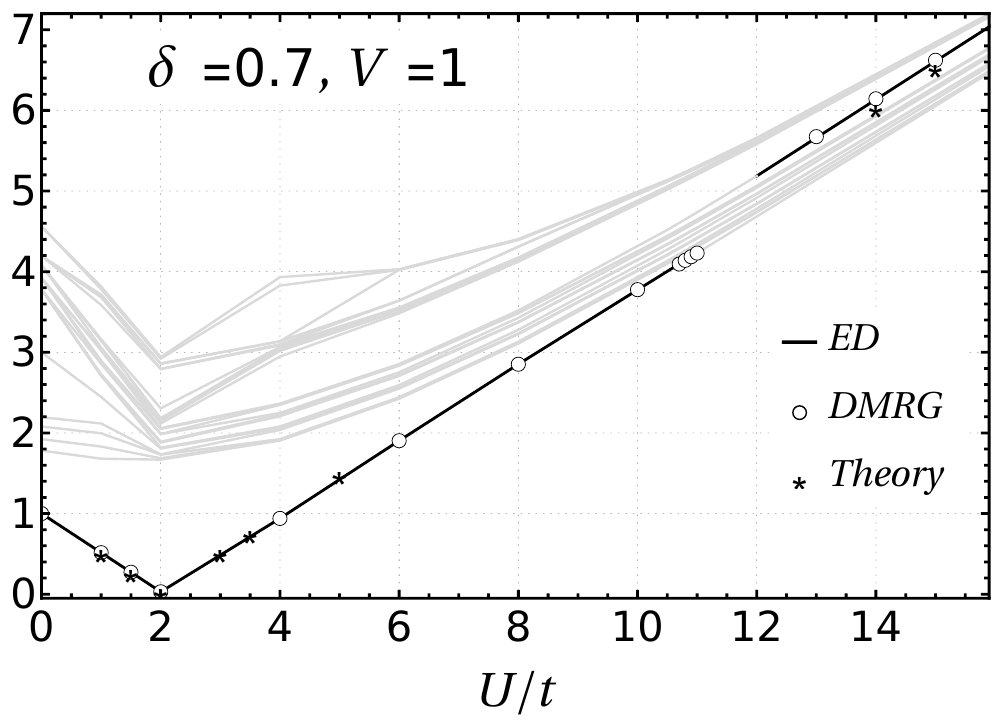}
		\caption{Energy eigenvalues of RMH model on an open chain of $10$ sites with $11$ electrons calculated using ED (exact diagonalization) are plotted here after subtracting the ground state energy at half-filling (i.e. $10$ electrons). 
		The edge state behaviour from ED (black line) compares very nicely with that from DMRG (empty circles) and theory (star).}
		\label{fig:L10_ED_N+1_excitation}
	\end{center}
\end{figure}

This transmigration can also be seen by staying in the $N_e=L+1$ sector itself. We use exact diagonalization to calculate a reasonable number ($\approx30$) of low excitation energies in this sector for $L=10$. After subtracting therefrom the ground state energy at half-filling, we plot these quasiparticle energies in Fig.~\ref{fig:L10_ED_N+1_excitation} for $\delta=0.7$, $V=1$, where the lowest excitation represents the edge state energy in the charge gap. As we increase $U$, the edge state energy (the black line) is clearly seen to transmigrate to the high energy gap for strong correlations; it compares reasonably with the edge state energies calculated from DMRG and the theory presented in the next section. 

The RMH model shows another interesting property besides transmigration, which is that the edge state energy invariably becomes zero at $U_c (\approx 2V)$. We noted this feature earlier in the discussion, but did not emphasise its novelty enough. See, for a non-zero $U$ or $V$, the edge states are generally found to occur at non-zero energies, which is the general expectation for interacting electrons. But surprisingly, here we always find the edge state energy to become zero at a certain $U_c$. 
Thus, in the RMH chain, we have a special situation where the `zero energy edge states' occur due to interaction! This is contrary to the general expectation. For a given $V$, the half-filled RMH model in the ground state is a band insulator for small $U$ and a Mott insulator for large $U$. The critical interaction, $U_c$, is where the change from band to Mott insulator seems to happen. So, on both sides of $U_c$, the edge state energy is non-zero, but at $U_c$, it becomes zero. This is similar to the interaction driven transition from the band to Mott insulator via metallic phase in ionic Hubbard model~\cite{garg2006prl}. But here the bulk gap never goes to zero; instead, the edge state energy vanishes. Notably, the transmigration occurs only when $U > U_c$. Below we develop a theory that describes these features nicely.

\section{\label{sec:RMHubbard_Theory} Theory of Charge Dynamics} 
To understand the numerically obtained edge state behaviour presented above, we investigate the properties of charge quasiparticles in RMH chain using Kumar representation. Following Ref.~\cite{bisht2024transmigration}, we write the RMH Hamiltonian, Eq.~\eqref{eq:RMHubbard}, in Kumar representation~\cite{kumar2008canonical} as

\begin{equation}
	\begin{split}
		&\Hhat = -\frac{t}{2}\sum_{l=1}^{L-1} \left[1+(-)^l \delta\right] \left\{\left(\fhat^\dagger_l \fhat^{ }_{l+1} + {\rm H.c.}\right)\left(1+\vec{\sigma}_l\cdot\vec{\sigma}_{l+1}\right) \right. \\
		& \left. + (-)^l \left(\fhat^\dagger_l \fhat^\dagger_{l+1} + {\rm H.c.}\right)\left(1-\vec{\sigma}_l\cdot\vec{\sigma}_{l+1}\right)  \right\} - V\sum_{l=1}^{L} (-)^l \sigma^z_l \fhat^\dag_l\fhat^{ }_l \\
		& - \frac{U}{2}\sum_{l=1}^L \fhat^\dag_l\fhat^{ }_l  + V\sum_{l=1}^{L} (-)^l \sigma^{z}_{l} + \frac{U}{4}L 
	\end{split} \label{eq:RMHkumar}
\end{equation}
where the spinless fermion operators, $\hat{f}_l$ and $\hat{f}^\dag_l$, describe the charge excitations of the system and the Pauli operators, $\vec{\sigma}_l$, account effectively for the spin dynamics. Kumar representation describes the electrons canonically in terms of these spinless fermions and Pauli operators~\cite{kumar2008canonical}. 

From Eq.~\eqref{eq:RMHkumar}, we obtain a simpler effective model for the charge dynamics by replacing the spin dependent operator, $\vec{\sigma}_l\cdot\vec{\sigma}_{l+1}$, by its average value in the bulk; this average value is denoted as $\rho_{1-}$ over the odd bonds and $\rho_{1+}$ for the even bonds. Moreover, $\sigma^{z}_{l}$ in the staggered potential term is replaced by the value $m_{l}$. For $m_l$, we take an idealized choice, $m_l=(-)^{l+1}$, motivated by the fact that the staggered potential tends the alternate sites to be doubly occupied or empty. We also tried involved estimations of $m_l$, but this simple choice works better qualitatively and also quantitatively. With these simplifications, we get the following effective model for the charge quasiparticles 
\begin{equation}
	\begin{split}
		\Hhat_c =& -\frac{t}{2}\sum_{l=1}^{L-1} \left[1+(-)^l \delta\right] \Big\{ \left[1+\rho^{ }_{1,(-)^l}\right] \fhat^\dagger_l \fhat^{ }_{l+1} + \\
		& (-)^l \left[1-\rho^{ }_{1,(-)^l}\right] \fhat^\dagger_l \fhat^\dagger_{l+1} + {\rm H.c.}\Big\} +  u\sum_{l=1}^L \fhat^\dag_l\fhat^{ }_l
	\end{split} \label{eq:Hc_final}
\end{equation}
whose energy dispersion in the bulk is given by 
\begin{equation}
	E_{k,\pm} = \sqrt{u^2 + |\alpha_k|^2 + |\beta_k|^2 \pm 2\sqrt{u^2|\beta_k|^2 + \{\Re(\alpha^*_k\beta_k)\}^2}}
	\label{eq:Ekpm}
\end{equation}
for $\alpha_{k} =  t[(1+\delta)(1-\rho_{1+})+(1-\delta)(1-\rho_{1-})e^{i2k}]/2$ and $\beta_{k} =  t[(1+\delta)(1+\rho_{1+})+(1-\delta)(1+\rho_{1-})e^{i2k}]/2$ for $k\in[-\frac{\pi}{2},\frac{\pi}{2}]$. Here $u=V-U/2$. Minimum value of $E_{k,-}$ gives the bulk charge gap. 

\begin{figure}[t]
	\begin{center}
		\includegraphics[width=\columnwidth]{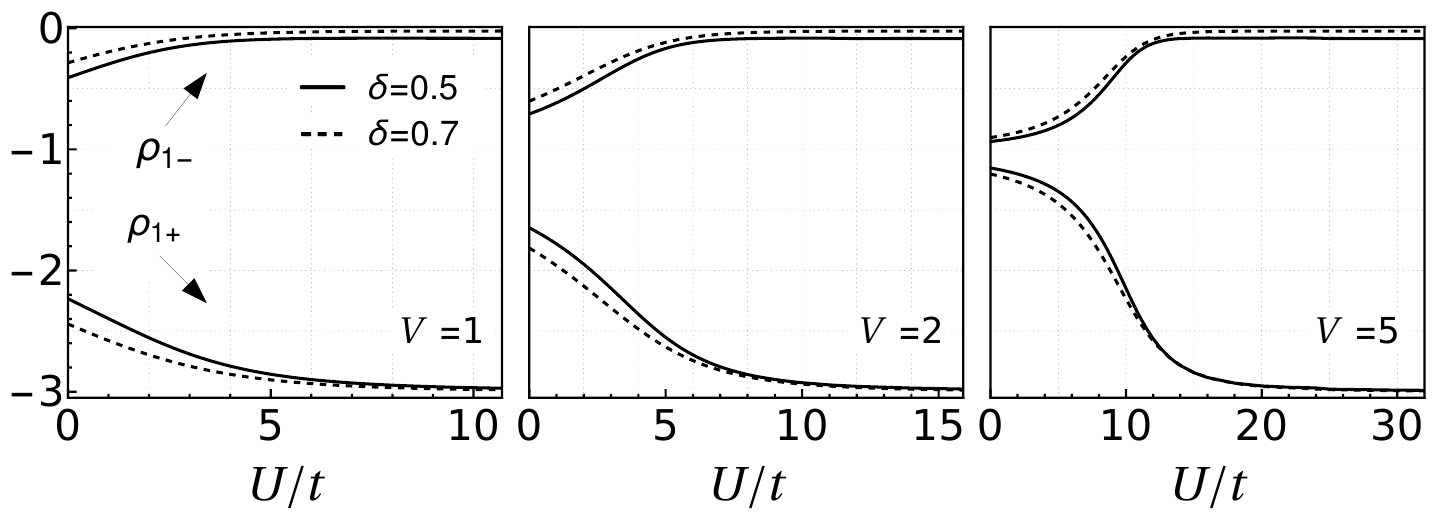}
		\caption{Parameters $\rho_{1\pm}$ required for the charge dynamics, Eq.~\eqref{eq:Hc_final}, calculated as a function of $U$ from DMRG in the ground state of the half-filled RMH model, Eq.~\eqref{eq:RMHubbard}.}
		\label{fig:rho1+-}
	\end{center}
\end{figure}

As in Ref.~\cite{bisht2024transmigration}, we calculate $\rho_{1\pm}$ in the ground state of the half-filled RMH model, Eq.~\eqref{eq:RMHubbard}, using the DMRG method. We calculated it for $L=60$, $100$ and $200$, and found the average values to not depend on the system size significantly. Figure~\ref{fig:rho1+-} shows the behaviour of $\rho_{1\pm}$ with $U$ for a few different values of $V$ and $\delta$. Note that a strong staggered potential needs a proportionately strong Hubbard interaction for $\rho_{1+}$ and $\rho_{1-}$ to get closer to the limiting values of $-3$ and $0$, respectively. We put in Eq.~\eqref{eq:Hc_final} the $\rho_{1\pm}$ thus obtained, and study the behaviour of charge quasiparticles for the RMH chain.

\begin{figure}[t]
	\begin{center}
		\includegraphics[width=\columnwidth]{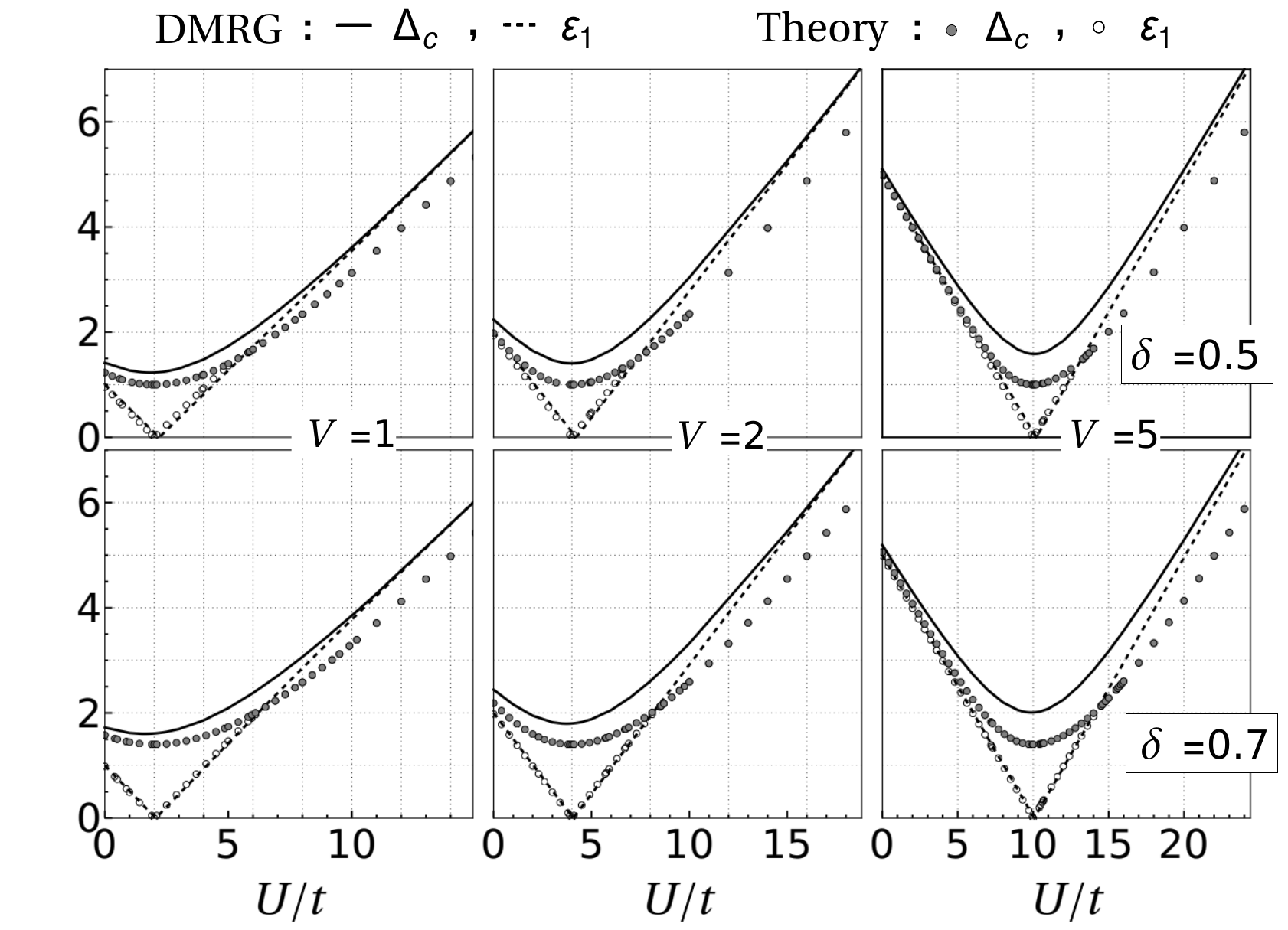}
		\caption{Comparing the charge gap ($\Delta_c$) and the edge state energy ($\varepsilon_1$) vs. $U$ from theory (filled and empty circles) and DMRG (solid and dashed lines) for a few fixed values of $V$ and $\delta$. Notably, the theory captures the cone-shaped variation of $\varepsilon_1$ with $U$ perfectly, but it underestimates $\Delta_c$.}
		\label{fig:delta=0.5_0.7_comparison_theory}
	\end{center}
\end{figure}

In Fig.~\ref{fig:delta=0.5_0.7_comparison_theory}, the empty circles represent the edge state energy obtained from our theory by solving Eq.~\eqref{eq:Hc_final} on open chain. Its quantitative match with the numerical data from DMRG (dashed line) is very good and the cone like feature of $\varepsilon_1$ vs. $U$ around $U_c$ is correctly produced by the theory. But the charge gap is underestimated by the theory (grey filled circles), although $\Delta_c$ vs. $U$ looks qualitatively similar to that from DMRG (black line). Consequently, we see a difference in the value of $U_{c,1}$ (where $\varepsilon_1$ touches $\Delta_c$) from theory and numerics. For example, for $\delta=0.7$ and $V=1$, we get $U_c\approx2.06~(2)$, $U_{c,1}\approx11~(6.51)$ and $U_{c,2}\approx13~(7.76)$ from the DMRG (theory). Inspite of these quantitative differences, there is a qualitative consistency in the physical behaviour seen from the numerics and this theory.

\begin{figure*}
	\begin{center}
		\includegraphics[width=\textwidth]{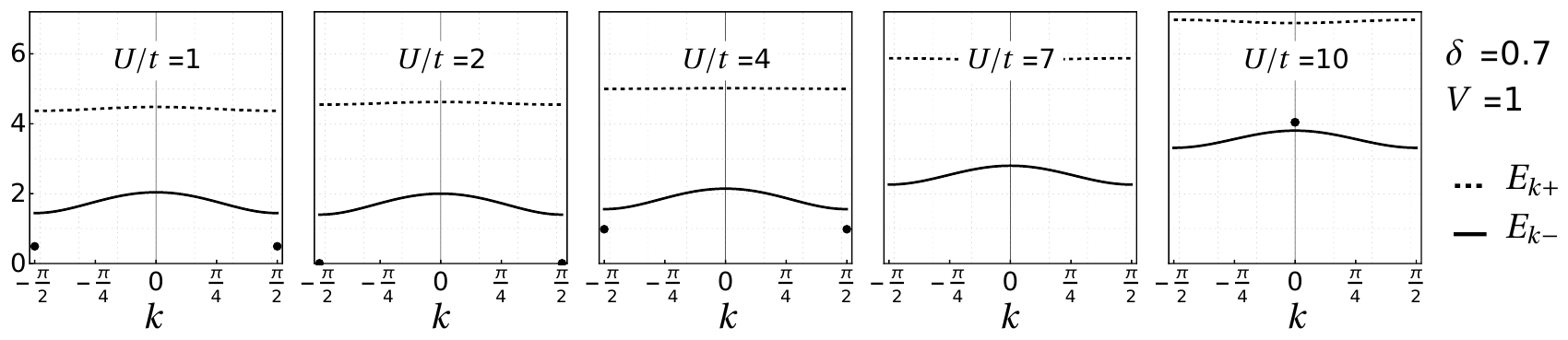}
		\caption{Evolution of the quasiparticle energy dispersion, Eq.~\eqref{eq:Ekpm}, with increasing $U$ for a fixed $\delta$ and $V$. Thick black dots represent the edge states. For $U=1$ and $2$, which are $< U_{c,1}$, the edge state occurs in the charge gap at $k=\pm\pi/2$. For $U=10$, which is  $>U_{c,2}$, the edge state is found at $k=0$ in the high energy gap. There is no edge state for $U=7$, which is in between $U_{c,1}$ and $U_{c,2}$. Notice the edge state at zero energy for $U=2$, i.e. $U_c$.} 
		\label{fig:Dispersion_edgestates_delta=0.7_V=1.0}
	\end{center}
\end{figure*}

In Fig.~\ref{fig:Dispersion_edgestates_delta=0.7_V=1.0}, we show the evolution of quasiparticle dispersion with $U$ for $\delta=0.7$ and $V=1$. Here the bulk dispersion, Eq.~\eqref{eq:Ekpm}, is presented together with the edge states (thick dots). We see different edge state behaviour in different regions across $U_c$, $U_{c,1}$ and $U_{c,2}$. For $U=1$, which is $< U_{c}$, the edge states in the charge gap lie at the zone boundary ($k=\pm\pi/2$) with a non-zero energy. At $U=2$, which is $U_c$, the edge states attain zero energy; the bulk always remains gapped. For $U=4$, which is $>U_c$, the edge state energy again becomes non-zero. After it merges into the bulk at $U_{c,1}$, no edge states are found; see here the plot for $U=7$. For $U>U_{c,2}$, e.g. $U=10$ in this figure, the edge states reappear in the high energy gap at the zone center ($k=0$). The transmigration of edge states from the charge gap to high energy gap, after having passed through zero energy, is quite evident here.

\begin{figure}[htbp]
	\begin{center}
		\includegraphics[width=\columnwidth]{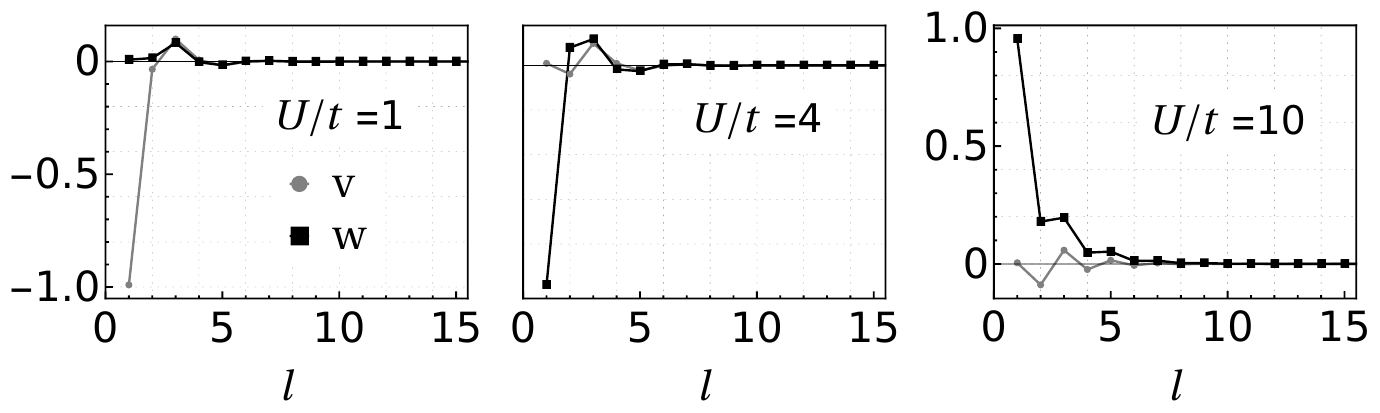} 
		\caption{Edge state wavefunction in three cases, $U=1 <U_{c}$, $U_{c}<U=4<U_{c,1}$, and $U=10>U_{c,2}$ for $\delta=0.7$ and $V=1$.}
		\label{fig:wavefunction_delta0p7_V1p0_new}
	\end{center}
\end{figure}

\begin{figure}[htbp]
	\begin{center}
		\includegraphics[width=\columnwidth]{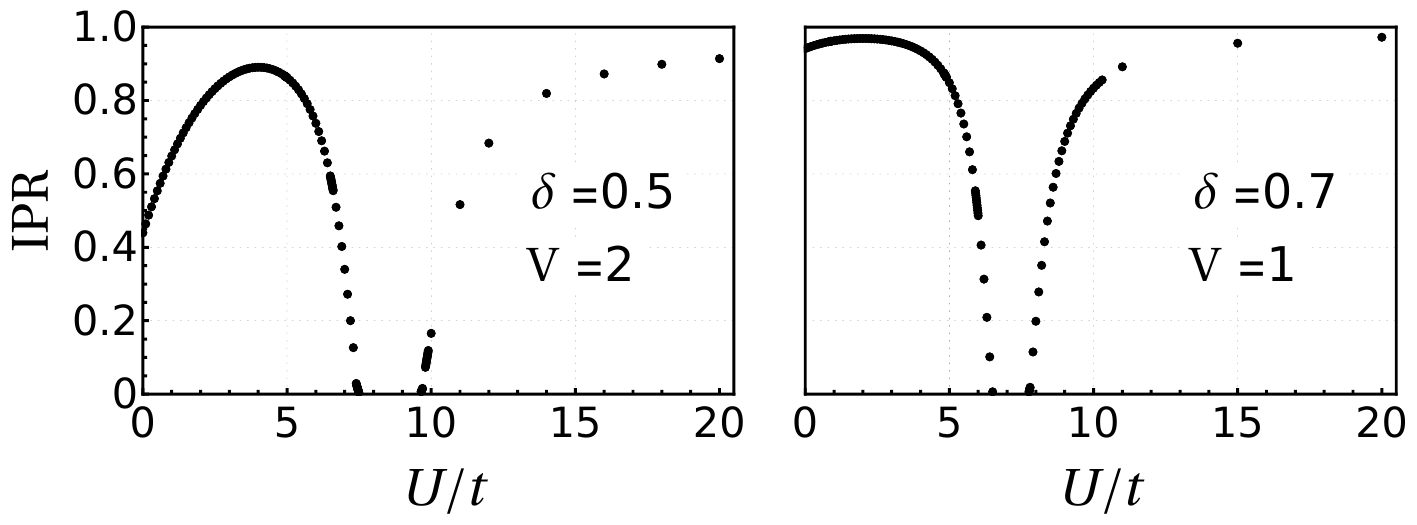}
		\caption{IPR (inverse participation ratio) vs. $U$. Here a non-zero value of IPR represents the presence of edge states.}
		\label{fig:IPR_0p5_0p7}
	\end{center}
\end{figure}

This theory lets us look into the quasiparticle edge states through their wavefunctions found by Bogoliubov diagonalization of $\hat{H_c}$ on open chain. Following Ref.~\cite{bisht2024transmigration}, we define the quasiparticle operator $\hat{\eta}$  
in terms of the spinless fermion $\hat{f}_l$ as $\hat{\eta}=\sum_{l=1}^{L} (v_l \hat{f}_{l} + w_l\hat{f}^\dagger_{l})$;  the coefficients $v_l$ and $w_l$ can be treated as the components of vectors $\mathbf{v}$ and $\mathbf{w}$, respectively. In Fig.~\ref{fig:wavefunction_delta0p7_V1p0_new}, we show the edge states computed in three different regions of $U$ for $\delta=0.7$ and $V=1$. The plot for $U=1$ corresponds to the region $U<U_{c}$; here $\mathbf{v}$ is peaked at $l=1$ and it rapidly dies off in the bulk. The plot for $U=4$ represents the region: $U_{c}<U<U_{c,1}$, wherein $\mathbf{w}$ is peaked at $l=1$. Note that $U_{c}$ is acting here like a chemical potential below which the wave function is particle like (with dominant $\mathbf{v}$) and above hole like (with dominant $\mathbf{w}$). In both these cases, the edge states lie in the physical charge gap at $k=\pi/2$. In the region $U_{c,1} < U < U_{c,2}$, we don't get any edge state. The plot for $U=10$ belongs to the case $U>U_{c,2}$, where the edge state occurs in the high energy gap and corresponds to $k=0$ or $\pi$. The  inverse participation ratio, IPR~$=\sum_{l=1}^{L}(|v_l|^4 + |w_l|^4)$, is a good identifier of the edge states. In Fig.~\ref{fig:IPR_0p5_0p7}, we plot IPR as a function of $U$ for $\delta=0.5$, $V=2$ and $\delta=0.7$, $V=1$. These plots clearly show the region, $U_{c,1}<U<U_{c,2}$, with no edge states. Notably, for $U<U_{c,1}$, the maxima of IPR is found to correspond to the zero energy edge states at $U_c$. 

\begin{figure}
	\begin{center}
		\includegraphics[width=\columnwidth]{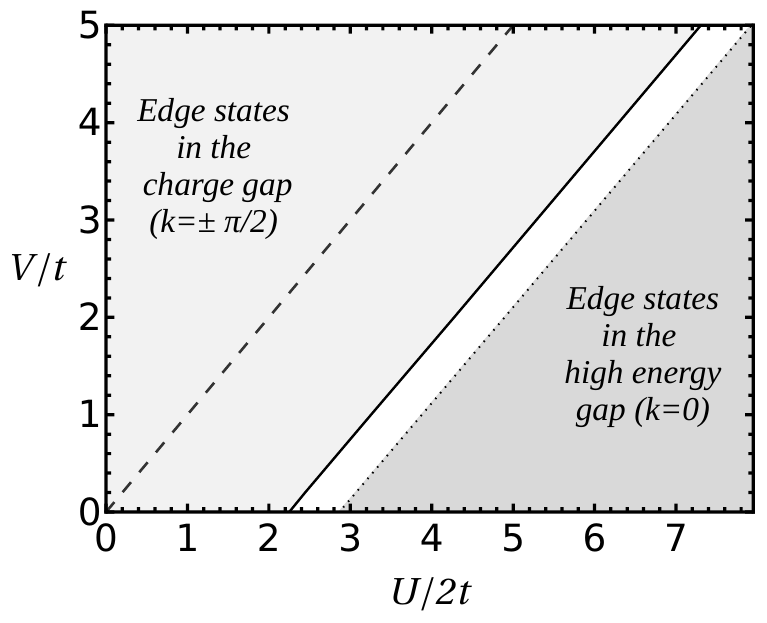}
		\caption{Phase diagram of the half-filled Rice-Mele-Hubbard chain for $\delta=0.7$, obtained from the theory of charge dynamics. The white region bounded by $U_{c,1}$ (solid line) and $U_{c,2}$ (dotted line) is the region where no edge states exist. In the light gray region for $U<U_{c,1}$, the edge states occur in the charge gap and the dashed line corresponds to $U_c$ where the edge states attain zero energy. In the gray region for $U>U_{c,2}$, the edge states occur in the high energy gap.}
		\label{fig:Phase_Diagram_delta=0.7}
	\end{center}
\end{figure}

Figure~\ref{fig:Phase_Diagram_delta=0.7} sums up the behaviour of edge states in RMH chain in the $U$-$V$ plane for a fixed $\delta$. Here there are three phases categorized according to the position of edge states in the energy spectrum. In first phase (light gray region), ranging from $U=0$ to $U_{c,1}$ (continuous line), the edge states are found in the physical charge gap. In the second phase (white region) from $U_{c,1}$ to $U_{c,2}$ (dotted line), no edge states exist; the width of this phase remains approximately the same for different values of the staggered potential for a given $\delta$. From $U_{c,2}$ onwards, the third phase (gray region) starts in which the edge states appear in high energy gap. In addition, here there is another very important feature in the first phase. It is that there exist zero energy edge states at $U_c$ (dashed line).

\section{\label{sec:summary} Summary}

We investigate the effect of Hubbard interaction on the edge states in Rice-Mele chain at half-filling and present some important results. The key findings of this study are as follows. For a fixed dimerization, $\delta$, and staggered potential, $V$, the edge state invariably transmigrates from the charge gap to the high energy gap by increasing the Hubbard interaction, $U$, with an intermediate phase devoid of edge states. We also find the edge state energy to always become zero at a critical interaction, $U_c\approx2V$, preceding  transmigration. It is interesting to see that the competition between $V$ and $U$ realizes zero energy edge states even for the interacting electrons.

\acknowledgements{J.B. acknowledges DST (India) for INSPIRE fellowship.}

\bibliography{manuscript_RMH}

\end{document}